\documentclass[%
 aip,
 amsmath,amssymb,
 reprint,%
]{revtex4-1}

\usepackage{graphicx}
\usepackage{dcolumn}
\usepackage{bm}

\usepackage[utf8]{inputenc}
\usepackage[T1]{fontenc}
\usepackage{mathptmx}
\usepackage{float}
\usepackage{subcaption}
\usepackage{url}

\begin{document}

\preprint{AIP/123-QED}

\title[Microwave Demonstration of Purcell Effect Enhanced Radiation Efficiency]{Microwave Demonstration of Purcell Effect Enhanced Radiation Efficiency}

\author{L. D. Stanfield}
    \email{lds211@exeter.ac.uk}

\author{A. W. Powell}

\author{S. A. R. Horsley}

\author{J. R. Sambles}

\author{A. P. Hibbins}

\affiliation{ 
Department of Physics and Astronomy, University of Exeter, Stocker Road, Exeter, EX4 4QL
}%

\date{\today}

\begin{abstract}
We experimentally demonstrate a Purcell effect-based design technique for improved impedance matching, and thus enhanced radiation efficiency from a small microwave emitter. Using an iterative process centred on comparing the phase of the radiated field of the emitter in air with that of the emitter in a dielectric environment, we optimise the structure of a  dielectric hemisphere above a ground plane surrounding a small monopolar microwave emitter in order to maximise its radiation efficiency. The optimised system shows very strong coupling between the emitter and two omnidirectional radiation modes at 2.00 GHz and 2.84 GHz, yielding Purcell enhancement factors of 8360 and 430 times increase respectively, and near perfect radiation efficiency.\\

\end{abstract}

\maketitle

\section{Introduction}

In 1946, Purcell predicted that the spontaneous emission rate of atoms are enhanced if positioned within an appropriately-sized resonant cavity \cite{Purcell46}. Manipulation of the physical environment about an emitter allows control of the local density of states (LDOS), a measure of the availability of electromagnetic states at a point in space \cite{Khosravi92, Barnes19}. The Purcell effect, as this came to be known, has been thoroughly investigated in quantum systems, including enhancement of spontaneous emission rates of atoms placed in a cavity \cite{Kleppner81} or near to an interface \cite{Drexhage68, Drexhage70}, quenching of emission rates \cite{Anger06}, and enhancement of the photoluminescence quantum yield in silicon nanocrystals \cite{Valenta19}. The Purcell effect has also been experimentally demonstrated in classical systems, for example, in acoustics with a Chinese gong over a rigid interface \cite{Langguth16}, and a simple electric and magnetic dipole above an interface \cite{Krasnok15}. However, there is still much to investigate as regards to applying the Purcell effect to real world classical systems as, thus far, studies have been limited to simple demonstrations.

In this paper, we present a novel approach for designing the local physical environment about an emitter to control its radiative behaviour, adapted from optical regime techniques to manipulate its LDOS. We then demonstrate the effectiveness of this technique through the design and experimental validation of a structured hemispherical dielectric placed on a ground plane about a rod emitter that yields highly efficient modes at remarkably low frequencies relative to the emitter size. This study gives a clear experimental demonstration of a Purcell-based design technique that can be applied to improve the performance of both nano-optical systems and microwave structures. 

\section{Theory}

By constructing a resonant cavity about an emitter, the emitter is able to weakly couple to and radiate into cavity modes. Depending on the spatial structure of the cavity mode excited, there can be enhancement or suppression of the LDOS available to the emitter.  The rate of emission relative to free space is described by the following Purcell formula:

\begin{equation}
F = \frac{\gamma}{\gamma_{\text{0}}} = 1 + \frac{6 \pi \epsilon_0^2}{|\textbf{d}|^2} \frac{1}{k^3} \text{Im} [\textbf{d}^{\ast} \cdot \textbf{E}_{\text{s}} (\textbf{r}_{\text{d}}) ].
\label{eq9}
\end{equation}

\noindent Equation \ref{eq9} is well known in quantum mechanics, and defines the Purcell factor, $F$, which is a ratio of the decay rate $\gamma$ of the quantum emitter embedded within a structure versus the decay rate in free space, $\gamma_0$. In the above formula, $\epsilon_0$ is the permittivity of free space, $\textbf{d}$ is the dipole moment, $k = \omega / c$ is  the free space wavenumber, $\omega$ is the operating frequency, $c$ is the speed of light in vacuum. $\textbf{E}_{\text{s}} (\textbf{r}_{\text{d}})$ is the net scattered field, evaluated at the position of the emitter $\textbf{r}_{\text{d}}$, that has been back-scattered by the surrounding dielectric environment  \cite{Krasnok15}.

Despite its origin, the application of Equation (\ref{eq9}) is not limited to quantum mechanics.  It is also applicable to classical systems, such as an acoustic speaker \cite{Landi18}, or a wire-based metamaterial \cite{Slobozhanyuk14}. It has also been shown by Krasnok et al\cite{Krasnok15} that various off-resonant antenna systems can be well described by Purcell effects. By employing the reciprocity theorem \cite{Mignuzzi19,Principles}, equation \ref{eq9} can be re-written as,

\begin{equation}
F = \frac{P_{\text{rad}}}{P_{0, \text{rad}}} = 1 + \frac{6 \pi \epsilon_0^2}{|\textbf{d}|^2} \frac{1}{k^3} \text{Im} \left[ \int_V d^3 \textbf{r} (\epsilon_r (\textbf{r}) - 1) f(\textbf{r}) \right].
\label{eq1}
\end{equation}

\noindent Equation \ref{eq1} describes the Purcell factor of a system in terms of radiated power from the emitter in free space, $P_{0, {\text{rad}}}$, and radiated power from the emitter within a dielectric structure, $P_{\text{rad}}$. Decay rate enhancement is analogous to radiated power enhancement \cite{Principles} under the assumption that the surrounding electromagnetic environment is lossless \cite{Krasnok15}. $f(\textbf{r})$ is defined as:

\begin{equation}
f(\textbf{r}) = \textbf{E}_{d \ast} (\textbf{r}) \cdot \textbf{E} (\textbf{r}).
\label{eq2}
\end{equation}

\noindent Equation \ref{eq2} allows the impact of radiation reaction in a structure to be quantified, where $\textbf{E}_{\text{d} \ast} (\textbf{r})$ is the electric field of the time-reversed dipole moment in free space\cite{Mignuzzi19}, and $\textbf{E} (\textbf{r})$ is the electric field of a dipole embedded within a structure. The electric field back-scattered by the surrounding dielectric environment can interfere either constructively or destructively at the position of the emitter. Constructive interference will result in stronger coupling between the emitter and the electromagnetic field, and destructive interference in weaker coupling. Equation \ref{eq2} can then be used to guide the iterative design of the dielectric structure, as discussed in Section \ref{Results}, where we apply the iterative design process theoretically presented by Mignuzzi et al\cite{Mignuzzi19} for nanoscale optical structures to the case of a real emitter operating in the microwave regime.\\

This investigation focuses on tailoring the immediate environment about a dipolar source in order enhance it's power emission and demonstrate a powerful Purcell enhancement for a small microwave emitter. However, the method described is very general and equation \ref{eq2} can be adapted for any radiation source whose electric field in free space can be described or simulated. The techniques used here therefore have the potential to act as a powerful tool for the design of devices like dielectric resonator antennas \cite{Leung93, Wong93, Powell21}, where more complex feeds often have advantages such as increasing the bandwidth or polarisation control \cite{McAllister83, Kranenburg88, Ghosh09}.

\section{Results and Discussion}
\label{Results}

\begin{figure}
\centering
\begin{subfigure}{1\linewidth}
   \centering
   \includegraphics[width=0.495\linewidth]{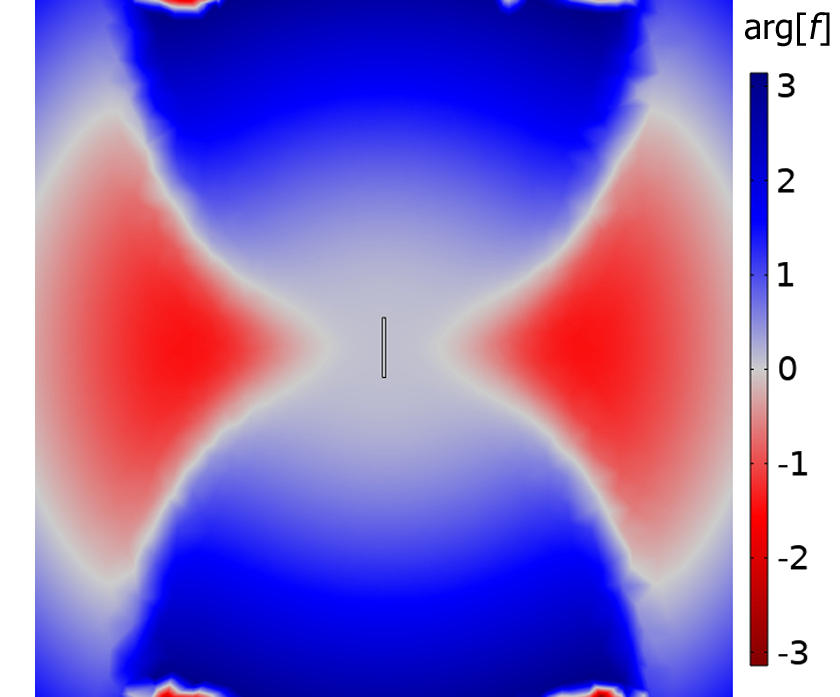}
   \includegraphics[width=0.495\linewidth]{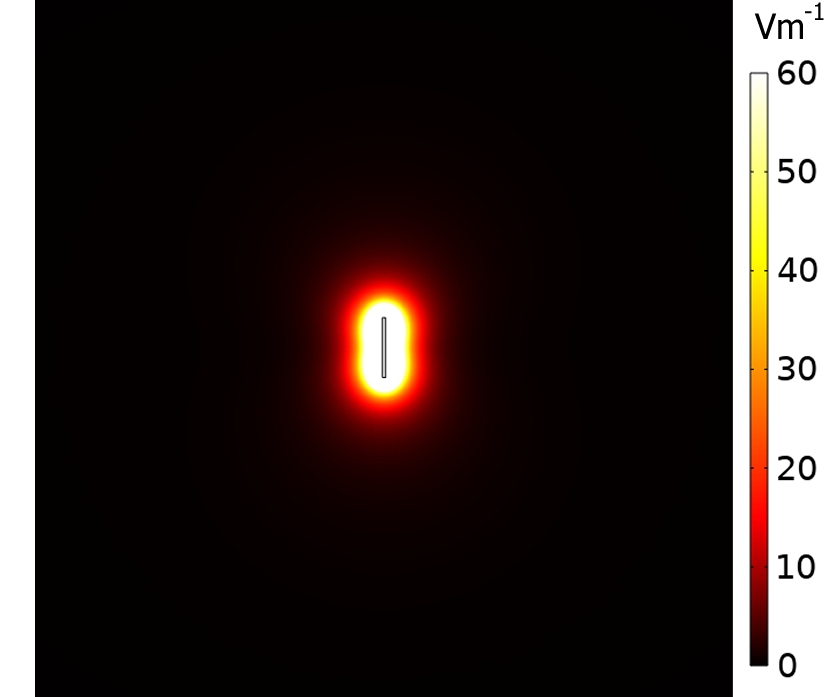}
   \caption{Dipole radiating in free space}
   \label{fig:dot1} 
\end{subfigure}
\\[\baselineskip]
\begin{subfigure}{1\linewidth}
   \centering
   \includegraphics[width=0.495\linewidth]{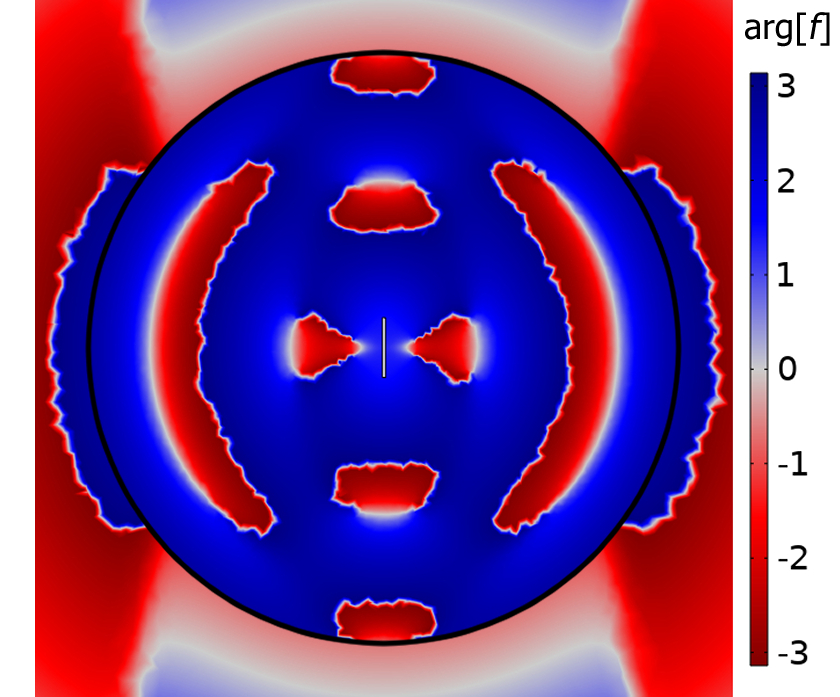}   \includegraphics[width=0.495\linewidth]{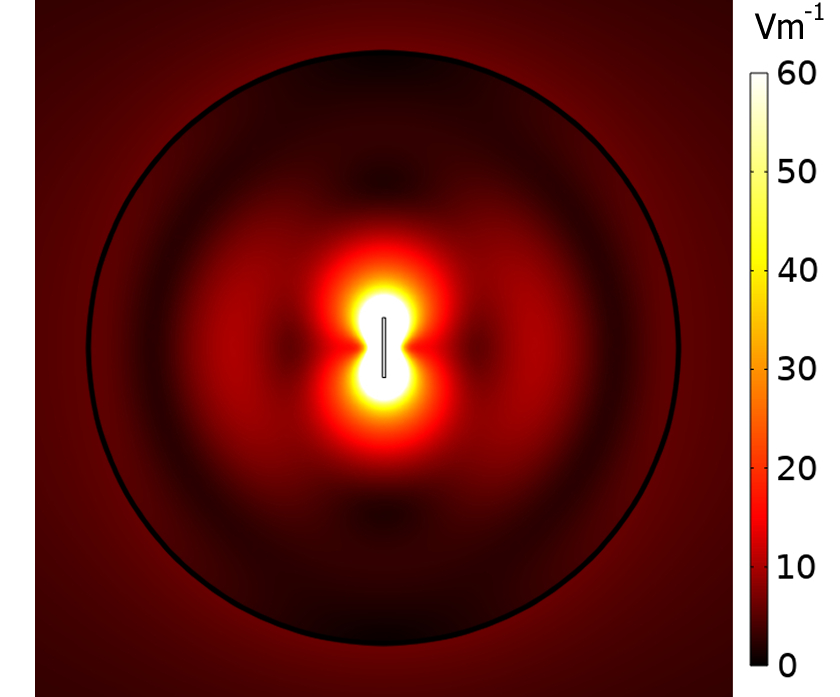}
   \caption{Dipole embedded in a dielectric sphere}
   \label{fig:dot2} 
\end{subfigure}
\\[\baselineskip]
\begin{subfigure}{0.495\linewidth}
   \centering
   \includegraphics[width=\linewidth]{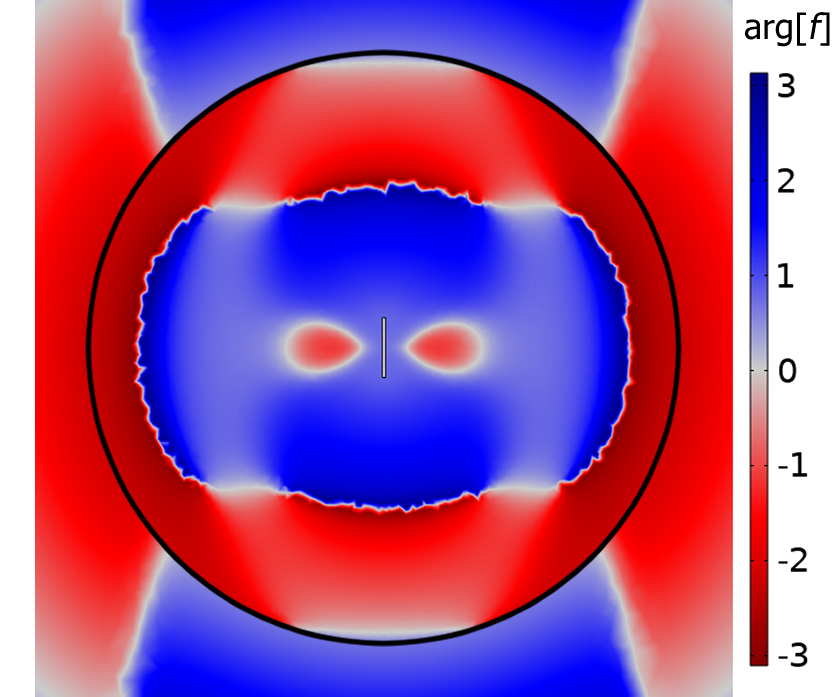}
   \caption{Dot product of (a) and (b)}
   \label{fig:dot3} 
\end{subfigure}
\hfill
\begin{subfigure}{0.495\linewidth}
   \centering
   \includegraphics[width=\linewidth]{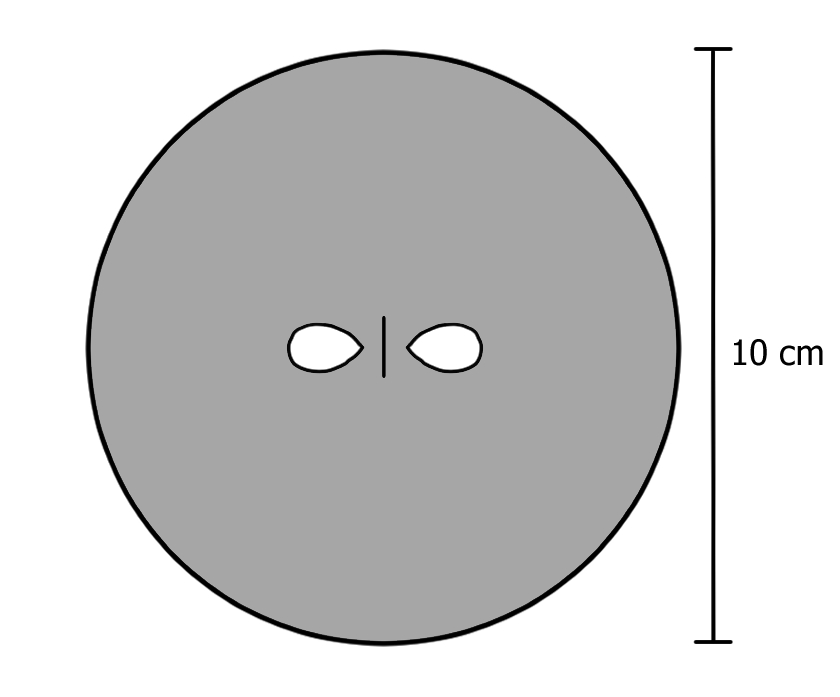}
   \caption{Iteration 1}
   \label{fig:dot4}
\end{subfigure}
\centering
\caption{(a) and (b) show the phase (left) and magnitude (right) of the numerically modelled electric fields at 2.00 GHz for a 1 cm dipole, radiating into free space, and that same dipole embedded in a 5 cm radius dielectric sphere of relative permittivity ($\epsilon_r = 12$) respectively. (c) displays arg[$f$], calculated using equation \ref{eq2}. (d) is the first iteration in the design process, informed by (c). Each image is displayed on the same scale.}
\end{figure}

By plotting arg[$f$] as a function of position, one can obtain a "phase map" comparing the electric field of a freely radiating dipole in free space, and the electric field of a dipole embedded within a dielectric structure, as shown in figure 1. A negative value corresponds to regions where a substrate would suppress the local density of states (LDOS) at the emitter and so the dielectric in these regions should be removed. 

\begin{figure*}
    \subfloat[Schematic of experimental set-up]{{\includegraphics[scale=0.75]{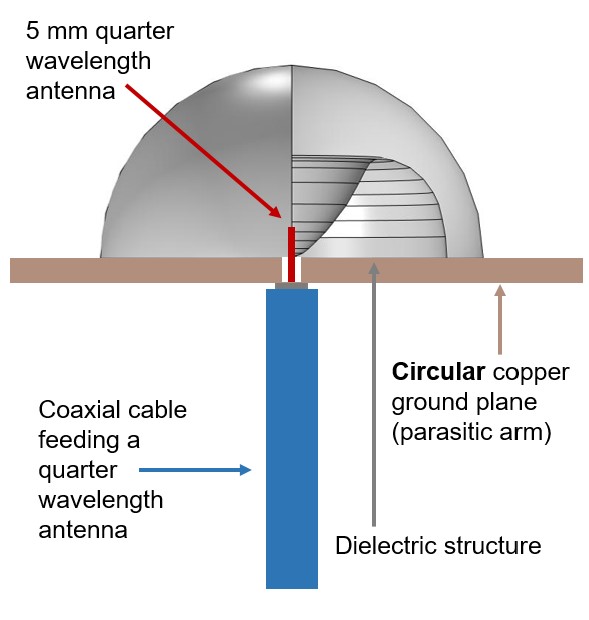} }}
    \qquad
    \subfloat[Photograph of experimental set-up]{{\includegraphics[scale=2.55]{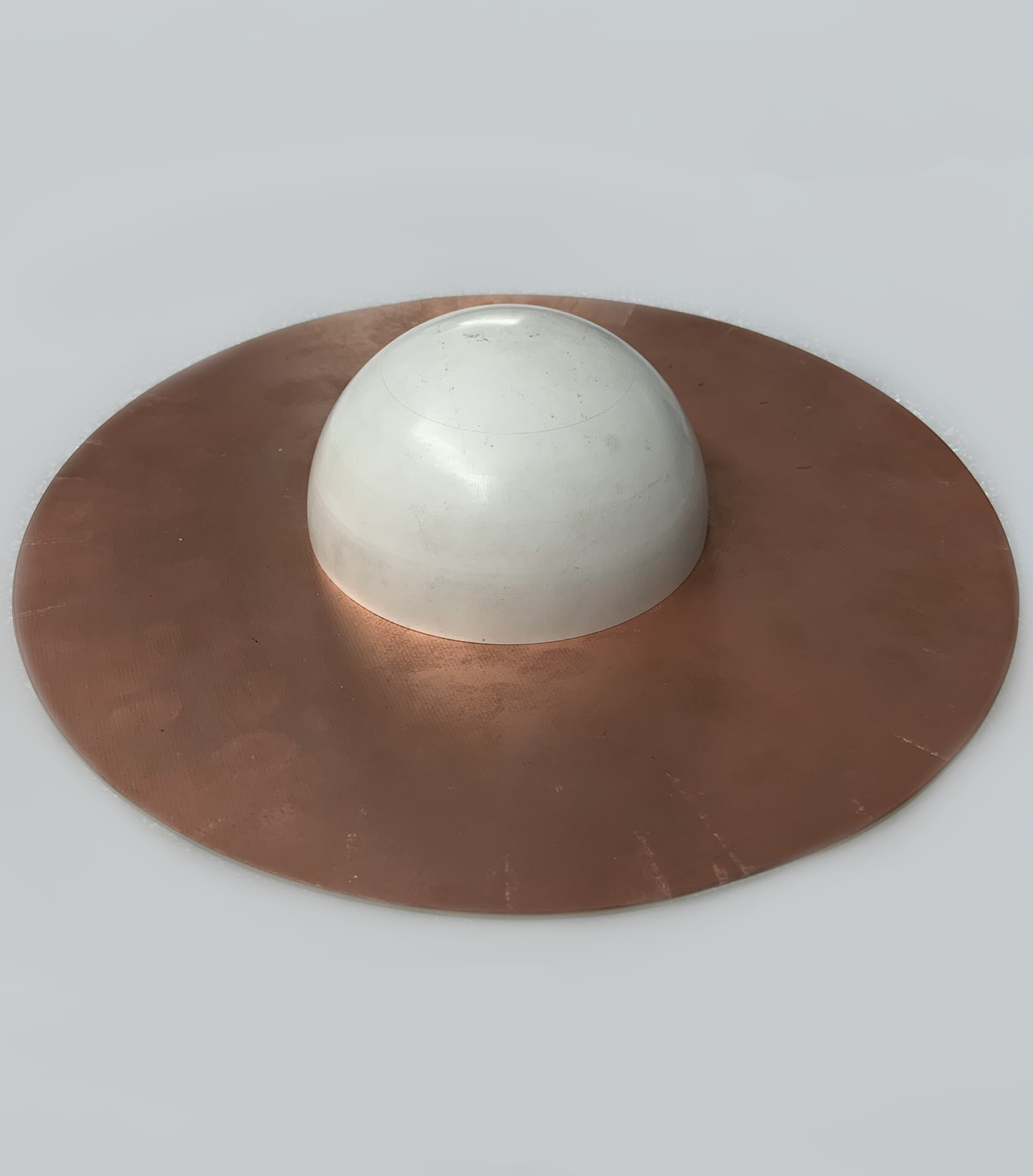} }}
    \caption{a) A schematic of the experimental set-up, shown on the x-z plane, where the 4.2 cm radius structured dielectric hemisphere of relative permittivity 12 is sat atop a 10 cm radius circular copper ground plane, with a coaxial-fed 0.5 cm grounded quarter-wavelength emitter feeding up through the center, and b) a photograph of said experimental set-up}
    \label{schematic}
\end{figure*}

The initial design process is conducted by embedding a 1 cm idealized dipole emitter of radius 0.05 cm within a 5 cm radius dielectric sphere - of relative permittivity 12, shown in figure 1 - in the x-z plane. The first step is to determine the phase difference between the electric fields of the free emitter and the embedded emitter, then to remove some dielectric volume, informed by the phase map, and monitor the effect of this removal on the reflection coefficient, defined as |S11|, calculated from a lumped port placed at the base of the feed using COMSOL. If more than one region is negative and should be removed, the changes closest to the emitter where the field magnitude is greatest are prioritised. Then a fresh "phase map" is produced using this structured sphere and this process of removal reshaping the hemisphere is repeated until the |S11| value on resonance converges to near 0, signifying near-perfect impedance matching. Finally, a 10 cm radius copper ground plane is introduced with a 0.5 cm grounded quarter-wavelength emitter fed up through the centre. A model with a structured hemisphere placed atop the ground plane and about the emitter is re-optimised. This new geometry is to allow for easier manufacture, and to better replicate what is feasible to create and experimentally test.

Once this iterative process was completed a structured hemisphere was machined from a block of PREPERM PPE1200 \cite{PrePerm}. This has a relative permittivity of $\epsilon_r = 12.0 \pm 0.5$, and a loss tangent of $\tan \delta = 0.001$, characterised at 2.4 GHz.

Figure \ref{schematic} presents the system set up. A 0.5 cm quarter wavelength emitter, with a 15 GHz fundamental resonance in free space, is inserted through a 10 cm radius ground plane, made from copper-coated circuit board. The 4.2 cm radius dielectric hemispherical structure is placed above the ground plane.

The system was then experimentally characterised using an Anritsu  MS46122B Vector Network Analyser (VNA), and a rotational table controlled by a Thorlabs APT Precision Motion Controller in an anechoic chamber. S11 measurements were taken over the frequency range 1.5 GHz to 3.5 GHz in 0.005 GHz increments. Further to determine the far-field radiation pattern, the structure underwent a full 360$^{\circ}$ rotation in 1$^{\circ}$ increments about the z-axis, parallel to the axis of the rod emitter, in an anechoic chamber, with response measured by a Flann Dual Polarized Horn Antenna (Model DP240) for frequency ranges of 1.9 GHz to 2.1 GHz and 2.75 GHz to 2.95 GHz in 0.01 GHz increments.\\

\begin{figure*}
\centering
\includegraphics[width=18.15cm]{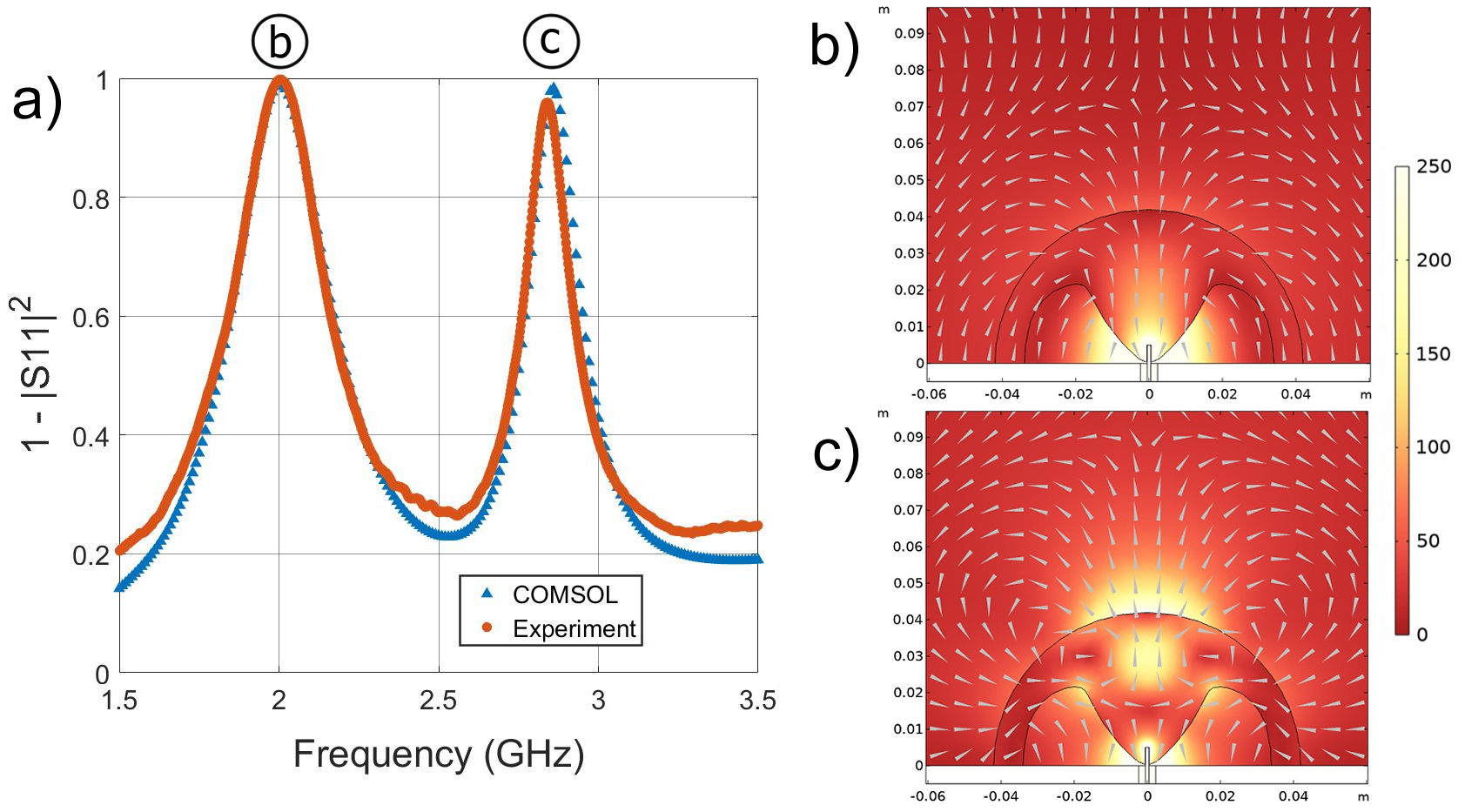} 
\caption{a) The ratio of input power to transmitted power as a function of frequency predicted in numerical modelling and observed in experiment, between 1.5 GHz and 3.5 GHz, and b) and c) the normalised electric field in the x-z plane with arrows indicating direction of field flow at 2.00 GHz and 2.86 GHz respectively, simulated via numerical modelling.}
\label{power}
\end{figure*}

Figure \ref{power} a) displays the computationally predicted and experimentally observed radiative behaviour of the system. The fundamental mode, for which the structured hemisphere was optimised, is numerically predicted and experimentally identified at 2.00 GHz, and is 99.0$\%$ efficient. In comparison, an ideal bare 0.5 cm quarter wavelength antenna inserted through a 10 cm radius circular copper ground plane would have a radiation efficiency of 0.012$\%$ at 2.00 GHz, and would radiate optimally at 15 GHz in free space - this is a 7.5 times reduction in operating frequency. The Purcell Factor at 2.00 GHz shows 8360 times enhancement of radiation efficiency versus the efficiency of the emitter radiating into free space, calculated using equation \ref{eq9}. The experiment with the dielectric hemisphere shows a resonance with a Q-factor of Q = 6.68 $\pm$ 0.10, while the numerical model predicts Q = 6.48. The minor difference in Q-factor can be explained by the raised baseline between peaks in the experimental data, which in the raw |S11| data, is 0.024 higher on average. The modelled electric field profile, shown in figure \ref{power} b), resembles the fundamental mode of a dipole in the far field.

\begin{figure*}
    \subfloat[Fundamental mode at 2.00 GHz]{{\includegraphics[width=8.56cm]{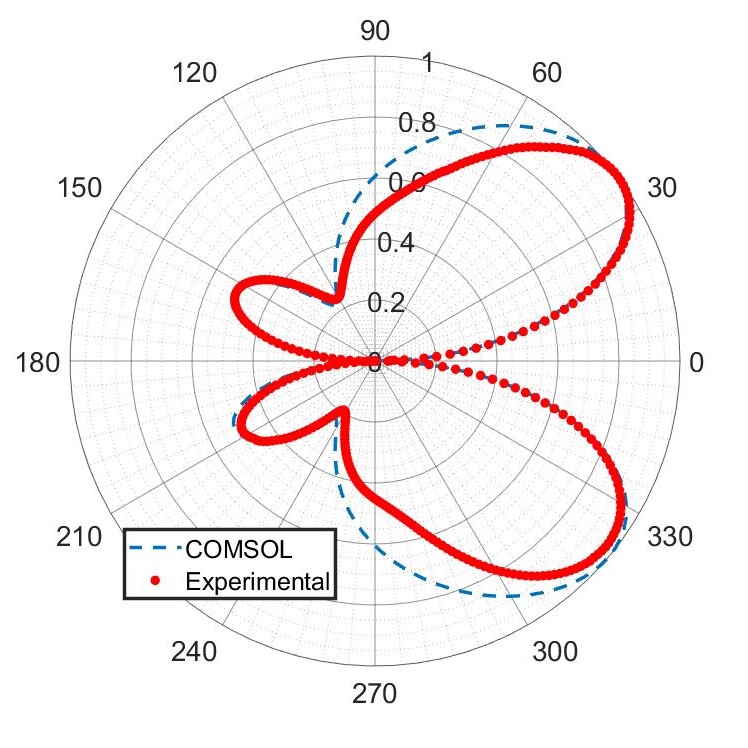} }}
    \qquad
    \subfloat[Higher order mode at 2.84 GHz]{{\includegraphics[width=8.56cm]{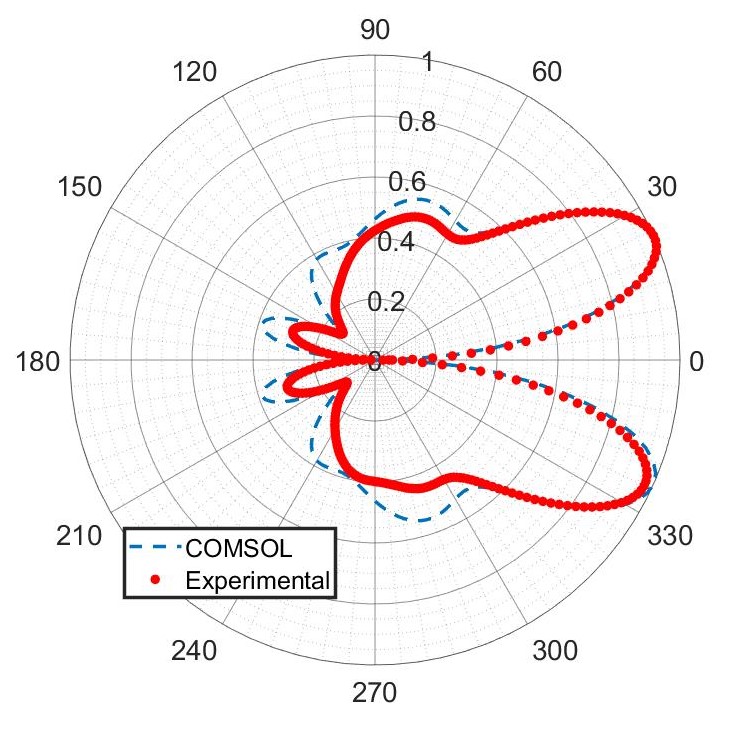} }}
    \caption{Polar plots of the normalised far-field radiation patterns in the x-y plane for the system at the identified (a) fundamental and (b) higher order mode, comparing computational and experimental data. The axis of the quarter-wavelength rod emitter is aligned with the 0$^{\circ}$ direction, and the ground plane is aligned along the 90$^{\circ}$-to-270$^{\circ}$ direction.}
    \label{graph1}
\end{figure*}

Surprisingly, with the same structure there is almost equally strong enhancement of a higher order mode from the same family. This is numerically predicted to be at 2.86 GHz and experimentally identified at 2.84 GHz, and is 98.3$\%$ efficient. The Purcell Factor at 2.84 GHz shows 430 times enhancement of radiation efficiency, calculated using equation \ref{eq9}. The experiment gives a resonance with a Q-factor of Q = 19.06 $\pm$ 0.286, whereas the numerical model predicts Q = 17.88. The modelled electric field profile, shown in figure \ref{power} c), is more complicated than that of the previous mode, but the whole structure resembles the linear third harmonic mode of a metal rod in the far field. Note that the structure was not optimised for coupling to this mode but because it is now so close in frequency to the fundamental then the system is still close to optimised. Both of these modes present within the radar S-band, defined as 2 to 4 GHz.\\

The radiation pattern for the 2.00 GHz fundamental mode, shown in figure \ref{graph1}(a), has two main lobes directed along $\pm 39^{\circ}$ about $0^{\circ}$, and, because of the finite size of the ground plane, two weak back lobes directed along 155$^{\circ}$ and 205$^{\circ}$. The radiation pattern for the 2.84 GHz third harmonic mode, shown in figure \ref{graph1}(b), has two main lobes directed along $\pm 26^{\circ}$ about $0^{\circ}$, and two back lobes directed along 162$^{\circ}$ and 198$^{\circ}$. The side features along approximately 90$^{\circ}$ and 270$^{\circ}$ correspond with radiation coming out perpendicular to the feed, and parallel to the ground plane. This is akin to the third harmonic mode of a simple rod antenna, where we also see minor side lobes perpendicular to the axis of the antenna.

In both radiation patterns, there is the expected null along the 0$^{\circ}$-to-180$^{\circ}$ direction, parallel with the orientation of the quarter-wavelength rod emitter. As the structure is axisymmetric about the axis of the quarter-wavelength emitter, the radiation patterns in figure \ref{graph1} are also axisymmetric, and form a diverging conical beams in the forwards direction.

The fundamental mode has an angular separation, and thus conical divergence of the main lobes of 78$^{\circ}$. The third harmonic mode is more tightly confined, with an angular spread of 52$^{\circ}$, and a FWHM main lobe width of 42$^{\circ}$.

\section{Conclusions}

This work experimentally demonstrates a novel technique based upon the Purcell effect for designing the local physical environment about a microwave source, in order to maximise its radiation efficiency at a specified non-resonant frequency. We have iteratively designed a dielectric structure surrounding a 0.5 cm quarter-wavelength rod emitter and have demonstrated very high radiation efficiency for a structure much smaller than its fundamental operational wavelength in air. Somewhat surprisingly the structure which has been optimised to operate at the fundamental resonance is also well optimised for the third harmonic. This offers thereby a dual-frequency system with two very efficient, well confined modes at very close and remarkably low frequencies, with both modes within S-band range. Our particular system displays a 2 GHz fundamental resonance with 99.0$\%$ efficiency, and a 2.84 GHz third harmonic resonance with 98.3$\%$ efficiency. Further optimisation would allow equally good efficiencies for higher order resonances.

 The technique could be utilised to optimise the design of dielectric resonator antennas with standard and non-standard feeds, and with further adaptation of the iterative technique could be used to enhance directivity, or other quantifiable measures of antenna behaviour. This would lend itself to many practical applications, namely within communications.\\

All data created during this research are openly available from the University of Exeter’s institutional repository.

\section{Acknowledgements}

L. D. Stanfield would like to thank James Capers and Cameron Gallagher for helpful conversations, and Ian Hooper for help in optimising the COMSOL modelling.

The authors acknowledge financial support from Leonardo Ltd UK and the Engineering and Physical Sciences Research Council (EPSRC) of the United Kingdom, via the EPSRC Centre for Doctoral Training in Metamaterials, under iCASE studentship grant EP/R511924/1.

\section{References}

\nocite{*}
\bibliography{aipsamp.bib}

\end{document}